\documentstyle[aps,epsfig]{revtex}

\makeatletter

\begin{document}

{\par\centering {\huge Angular correlation theory for double photoionization in a rare gas atom :
ionization by polarized photons}\huge \par}

\vspace{0.5in}

{\par\centering {\large  Chiranjib Sur and Dipankar Chattarji}\large \par}

{\par\centering \emph {Department of Physics, Visva-Bharati, Santiniketan 731 235, INDIA} \par}

\vspace{0.3in}
{\small This is a sequel to an earlier article on the theory of angular correlation
for double photoionization. Here we consider the two-step double photoionization
of a rare gas atom under the influence of a polarized photon beam described
by appropriate Stokes parameters. Cylindrical mirror analyzers (CMA) are used
to detect the outgoing electrons. Theoretical values of the correlation function
are obtained for linearly polarized light. Two different situations are handled.
Once, the value of the correlation function is obtained keeping the photo-electron
in a fixed direction. In the other case the direction of the Auger electron
is kept fixed. Comparison with experiments on xenon shows excellent agreement
for the case of \( 4d_{5/2} \) photoionization followed by a subsequent \( N_{5}-O_{23}O_{23}\, ^{1}S_{0} \)
Auger decay for a linearly polarized incident photon of energy \( 94.5eV \)
{[}J. Phys.B, \textbf{26}, 1141 (1993){]}.{\small \par}
\vspace{0.2in}

\vspace{0.2in}

\section{Introduction}

In an earlier paper {[}\ref{we2}{]} we considered the double photoionization
(DPI) of a rare gas atom under the influence of a unpolarized photon. The atom
was taken to be in a randomly oriented \( ^{1}S_{0} \) state. We considered
the angular correlation between the two successively emitted electrons, their
emissions being adequately separated in time {[}\ref{we1}{]}. Using a statistical
theory we obtained good agreement with the experimental results of K\( \ddot{a} \)mmerling
and Schmidt {[}\ref{schmidt}{]}.

In the present paper we take the incident photon beam to be polarized. The rare
gas atoms receiving the photon beam no longer remain randomly oriented, but
become aligned. If a photon of adequate energy is absorbed by an atom, a photo-electron
is emitted from one of its inner shells, leaving the atom singly ionized. This
ion subsequently de-excites by emitting an Auger electron {[}\ref{dc}{]} from
one of its outer shells. We are left with a doubly ionized atom and two electrons
in the continuum.The double photoionization process described above therefore
amounts to

\begin{equation}
\label{1}
h\nu +\textrm{A }\longrightarrow \textrm{A}^{+}+e^{-}_{1}\longrightarrow \textrm{A}^{++}+e_{1}^{-}+e_{2}^{-}.
\end{equation}
 As in reference {[}\ref{we2}{]} we denote the initial state (photon+atom)
by the set of quantum numbers \( (J_{a}M_{a}\alpha _{a}) \), or by virtual
quantum numbers \( (J^{\prime }_{a}M^{\prime }_{a}\alpha ^{\prime }_{a}) \),
keeping in mind possible interaction with other atoms and electrons. \( (J_{a},M_{a}) \)
or \( (J^{\prime }_{a}M^{\prime }_{a}) \) are angular momentum quantum numbers,
and \( \alpha _{a} \),\( \alpha ^{\prime }_{a} \) stand for the set of remaining
quantum numbers. Similarly for the intermediate and final states. The polarization
properties of the photon beam are described by appropriate Stokes parameters
\( S_{1},S_{2} \) and \( S_{3} \){[}\ref{born}{]}.

\section{DPI by polarized photons : }

We proceed by calculating the density matrix {[}\ref{blum}{]} and the angular
correlation function, which is the expectation value of the efficiency operator
for the detection of electrons. The density matrix of the initial state equals
the product of the density matrix of the intermediate singly ionized atom and
the density matrix of the photo-electron. Similarly, the density matrix of the
singly ionized atom can be written as the product of the density matrices of
the doubly ionized atom and the Auger electron. 

Using the Wigner-Eckart theorem, the matrix element of the density operator
for the initial atomic state can be expressed as {[}\ref{we2}{]}

\begin{equation}
\label{2}
\left\langle J_{a}M_{a}\alpha _{a}\right| \rho \left| J^{\prime }_{a}M^{\prime }_{a}\alpha ^{\prime }_{a}\right\rangle =\sum _{k_{a}\kappa _{a}}(-1)^{J^{\prime }_{a}-M^{\prime }_{a}}C_{M_{a}M^{\prime }_{a}\kappa _{a}}^{J_{a}J^{\prime }_{a}k_{a}}\rho _{k_{a}\kappa _{a}}(J_{a}\alpha _{a},J^{\prime }_{a}\alpha ^{\prime }_{a})\, .
\end{equation}
 Here the statistical tensor \( \rho _{k_{a}\kappa _{a}} \) is an irreducible
tensor of rank \( k_{a} \), which transforms according to the \( (2k_{a}+1) \)
dimensional irreducible representation \( D^{k_{a}} \) of the rotation group.
In Eq.(\ref{2}) \( C_{M_{a}M_{a}\kappa _{a}}^{J_{a}J^{\prime }_{a}k_{a}} \)
is a Clebsch-Gordan coefficient satisfying the triangle rule \( \mathbf{k}_{\mathbf{a}}=\mathbf{J}_{\mathbf{a}}+\mathbf{J}^{\prime }_{\mathbf{a}} \)
and \( \kappa _{a} \) is the projection of \( k_{a} \). Using the unitarity
property of Clebsch-Gordan coefficients we get

\begin{equation}
\label{3}
\rho _{k_{a}\kappa _{a}}(J_{a}\alpha _{a},J^{\prime }_{a}\alpha ^{\prime }_{a})=\sum _{M_{a}M^{\prime }_{a}}(-1)^{M_{a}-M^{\prime }_{a}}C_{M_{a}M_{a}\kappa _{a}}^{J_{a}J^{\prime }_{a}k_{a}}\left\langle J_{a}M_{a}\alpha _{a}\right| \rho \left| J^{\prime }_{a}M^{\prime }_{a}\alpha ^{\prime }_{a}\right\rangle \, .
\end{equation}
 We assume that the initial state is formed after the randomly oriented rare
gas atom absorbs a photon. Then the density matrix of the initial state becomes

\begin{equation}
\label{4}
\begin{array}{cc}
\rho _{k_{a}\kappa _{a}}(J_{a}\alpha _{a},J^{\prime }_{a}\alpha ^{\prime }_{a})= & 3(2J_{0}+1)\sum _{k_{0}\kappa _{0}k_{\gamma }\kappa _{\gamma }}\sqrt{2k_{0}+1}\sqrt{2k_{\gamma }+1}C_{\kappa _{0}\kappa _{\gamma }\kappa _{a}}^{k_{0}k_{\gamma }k_{a}}\\
 & \times \left\{ \begin{array}{ccc}
J_{0} & J^{\prime }_{0} & k_{0}\\
1 & 1 & k_{\gamma }\\
J_{a} & J_{a} & k_{a}
\end{array}\right\} \rho _{k_{0}\kappa _{0}}(J_{0},J_{0})\rho _{k_{\gamma }\kappa _{\gamma }}^{\gamma }(1,1)\, .
\end{array}
\end{equation}
 This equation satisfies the triangle rule \( \mathbf{k}_{0}=\mathbf{J}_{0}+\mathbf{J}^{\prime }_{0} \),
where \( J_{0} \) and \( J^{\prime }_{0} \) are the angular momentum quantum
numbers of the randomly oriented atom before absorption of the photon and its
virtual counterpart respectively. Here \( \rho _{k_{0}\kappa _{0}}(J_{0},J_{0}) \)
represents the density matrix of the randomly oriented atom and can be expressed
as

\begin{equation}
\label{5}
\rho _{k_{0}\kappa _{0}}(J_{0},J_{0})=\frac{1}{\sqrt{2J_{0}+1}}\delta _{k_{0}0}\delta _{\kappa _{0}0}\left\langle J_{b}\right\Vert j_{1}\left\Vert J_{a}\right\rangle \left\langle J_{b}\right\Vert j^{\prime }_{1}\left\Vert J_{a}\right\rangle ^{\star }\left\langle J_{c}\right\Vert j_{2}\left\Vert J_{b}\right\rangle \left\langle J_{c}\right\Vert j^{\prime }_{2}\left\Vert J_{b}\right\rangle ^{\star }.
\end{equation}
 Here the symbol \( \left\langle \right\Vert \left\Vert \right\rangle  \) stands
for a reduced matrix element.

\noindent In Eq. (\ref{4}) the expression \( \rho _{k_{\gamma }\kappa _{\gamma }}^{\gamma }(1,1) \)
represents the density matrix of the photon with its polarization properties.
Its elements are

\begin{equation}
\label{6}
\begin{array}{ccc}
\rho ^{\gamma }_{00}=\frac{1}{\sqrt{3}} & \rho ^{\gamma }_{10}=\frac{S_{3}}{\sqrt{3}} & \rho ^{\gamma }_{20}=\frac{1}{\sqrt{6}}\\
\rho ^{\gamma }_{1\pm 1}=0 & \rho ^{\gamma }_{2\pm 1}=0 & \rho ^{\gamma }_{2\pm 2}=-\frac{1}{2}(S_{1}\mp iS_{2})\, .
\end{array}
\end{equation}
 \( S_{1},S_{2} \) and \( S_{3} \) are the Stokes parameters {[}\ref{born}{]}
describing the polarization of the photon.

\noindent Then Eq.(\ref{4}) yields

\begin{equation}
\label{7}
\begin{array}{ccc}
\rho _{k_{a}\kappa _{a}}(J_{a}J^{\prime }_{a}) & = & 3(-1)^{J_{a}+k_{a}+1}\left\{ \begin{array}{ccc}
J_{a} & 1 & 0\\
1 & J_{a} & k_{a}
\end{array}\right\} \rho _{k_{a}\kappa _{a}}^{\gamma }(1,1)\\
 &  & \times \left\langle J_{b}\right\Vert j_{1}\left\Vert J_{a}\right\rangle \left\langle J_{b}\right\Vert j^{\prime }_{1}\left\Vert J_{a}\right\rangle ^{\star }\left\langle J_{c}\right\Vert j_{2}\left\Vert J_{b}\right\rangle \left\langle J_{c}\right\Vert j^{\prime }_{2}\left\Vert J_{b}\right\rangle ^{\star }\\
 & = & \frac{3(-1)^{J_{a}}}{\sqrt{2J_{a}+1}\sqrt{2J^{\prime }_{a}+1}}\rho _{k_{a}\kappa _{a}}^{\gamma }(1,1)\left\langle J_{b}\right\Vert j_{1}\left\Vert J_{a}\right\rangle \left\langle J_{b}\right\Vert j^{\prime }_{1}\left\Vert J_{a}\right\rangle ^{\star }\left\langle J_{c}\right\Vert j_{2}\left\Vert J_{b}\right\rangle \left\langle J_{c}\right\Vert j^{\prime }_{2}\left\Vert J_{b}\right\rangle ^{\star }.
\end{array}
\end{equation}

We define the angular correlation function as the expectation value of the efficiency
operator {[}\ref{we2}{]}. Following the same notation as in reference {[}\ref{we2}{]}
we can write it as

\begin{equation}
\label{8}
\overline{\varepsilon }=\sum _{J_{a}J^{\prime }_{a}\alpha _{a}\alpha ^{\prime }_{a}k_{a}\kappa _{a}}\rho _{k_{a}\kappa _{a}}(J_{a}\alpha _{a},J^{\prime }_{a}\alpha ^{\prime }_{a})\varepsilon ^{\star }_{k_{a}\kappa _{a}}(J_{a}\alpha _{a},J^{\prime }_{a}\alpha ^{\prime }_{a})\, .
\end{equation}
 Some simplification gives

\begin{equation}
\label{9}
\begin{array}{cc}
\overline{\varepsilon } & =\sum \rho _{k_{a}\kappa _{a}}(J_{a},J^{\prime }_{a})\varepsilon ^{\star }_{k_{c}\kappa _{c}}(J_{c},J^{\prime }_{c})\varepsilon ^{\star }_{k_{1}\kappa _{1}}(J_{1},J^{\prime }_{1})\varepsilon ^{\star }_{k_{2}\kappa _{2}}(J_{2},J^{\prime }_{2})\\
 & \times C_{\kappa _{b}\kappa _{1}\kappa _{a}}^{k_{b}k_{1}k_{a}}C_{\kappa _{c}\kappa _{2}\kappa _{b}}^{k_{c}k_{2}k_{b}}\sqrt{2J_{a}+1}\sqrt{2J^{\prime }_{a}+1}\sqrt{2k_{b}+1}\sqrt{2k_{1}+1}\\
 & \times \sqrt{2J_{b}+1}\sqrt{2J^{\prime }_{b}+1}\sqrt{2k_{c}+1}\sqrt{2k_{2}+1}\\
 & \times \left\{ \begin{array}{ccc}
J_{c} & j_{2} & J_{b}\\
J^{\prime }_{c} & j^{\prime }_{2} & J^{\prime }_{b}\\
k_{c} & k_{2} & k_{b}
\end{array}\right\} \left\{ \begin{array}{ccc}
J_{b} & j_{1} & J_{a}\\
J^{\prime }_{b} & j^{\prime }_{1} & J^{\prime }_{a}\\
k_{b} & k_{1} & k_{a}
\end{array}\right\} \, ,
\end{array}
\end{equation}
 where the summation extends over \( J_{a},J^{\prime }_{a},J_{b},J^{\prime }_{b},J_{c},J^{\prime }_{c},j_{1},j^{\prime }_{1},j_{2},j^{\prime }_{2},k_{a},\kappa _{a},k_{c},\kappa _{c},k_{1},\kappa _{1},k_{2} \)
and \( \kappa _{2} \).

In Eq.(\ref{9}) \( \varepsilon ^{\star }_{k_{i}\kappa _{i}}(j_{i},j^{\prime }_{i}) \)
is the efficiency tensor component for detection of the \( i \)th electron.
Here \( i=1 \) corresponds to the photo-electron, and \( i=2 \) to the Auger
electron. In DPI experiments the detectors usually used are cylindrical mirror
analyzers (CMA) {[}\ref{cma}{]} which have cylindrical symmetry with respect
to the axis of the detector. Details of the choice of detectors are given in
reference{[}\ref{we3}{]}. The efficiency tensor component now becomes

\begin{equation}
\label{10}
\varepsilon _{k_{i}\kappa _{i}}^{\star }(j_{i}j^{\prime }_{i})=\sum _{\kappa ^{\prime }_{i}}z_{k_{i}}(i)c_{k_{i}\kappa ^{\prime }_{i}}(j_{i}j^{\prime }_{i})D_{\kappa ^{\prime }_{i}\kappa _{i}}^{k_{i}}(\Re _{i})\, .
\end{equation}

Since the residual doubly ionized state is unobserved, the corresponding quantum
numbers are averaged over. This gives

\begin{equation}
\label{11}
\varepsilon _{k_{c}\kappa _{c}}^{\star }(J_{c}J^{\prime }_{c})=\sqrt{2J_{c}+1}\delta _{k_{c}0}\delta _{\kappa _{c}0}\delta _{J_{c}J^{\prime }_{c}}\, .
\end{equation}
 Then Eq.(\ref{9}) becomes

\begin{equation}
\label{12}
\begin{array}{cc}
\overline{\varepsilon } & =\sum \rho _{k_{a}\kappa _{a}}(J_{a},J^{\prime }_{a})\sqrt{2J_{c}+1}C_{\kappa _{b}\kappa _{1}\kappa _{a}}^{k_{b}k_{1}k_{a}}C_{0\kappa _{2}\kappa _{b}}^{0k_{2}k_{b}}\sqrt{2J_{a}+1}\sqrt{2J^{\prime }_{a}+1}\\
 & \times \sqrt{2k_{b}+1}\sqrt{2k_{1}+1}\sqrt{2J_{b}+1}\sqrt{2J^{\prime }_{b}+1}\sqrt{2k_{c}+1}\sqrt{2k_{2}+1}\\
 & \times \left\{ \begin{array}{ccc}
J_{c} & j_{2} & J_{b}\\
J^{\prime }_{c} & j^{\prime }_{2} & J^{\prime }_{b}\\
0 & k_{2} & k_{b}
\end{array}\right\} \left\{ \begin{array}{ccc}
J_{b} & j_{1} & J_{a}\\
J^{\prime }_{b} & j^{\prime }_{1} & J^{\prime }_{a}\\
k_{b} & k_{1} & k_{a}
\end{array}\right\} \\
 & \times z_{k_{1}}(1)c_{k_{1}\kappa ^{\prime }_{1}}(j_{1}j^{\prime }_{1})z_{k_{2}}(2)c_{k_{2}\kappa ^{\prime }_{2}}(j_{2}j^{\prime }_{2})D_{\kappa ^{\prime }_{1}\kappa _{1}}^{k_{1}}(\Re _{1})D_{\kappa ^{\prime }_{2}\kappa _{2}}^{k_{2}}(\Re _{2})\, .
\end{array}
\end{equation}
 Here we have used the relation

\begin{equation}
\label{13}
D_{\kappa ^{\prime }_{1}\kappa _{1}}^{k_{1}}(\Re _{1})D_{\kappa ^{\prime }_{2}\kappa _{2}}^{k_{2}}(\Re _{2})=\sum _{k}C_{\kappa _{1}\kappa _{2}\kappa }^{k_{1}k_{2}k}C_{\kappa ^{\prime }_{1}\kappa ^{\prime }_{2}\kappa ^{\prime }}^{k_{1}k_{2}k}D_{\kappa \kappa ^{\prime }}^{k}(\Re )
\end{equation}
 to get the actual angular dependence of the angular correlation function. In
Eq.(\ref{13}) the Euler rotation \( \Re =(\beta _{1}\theta \beta _{2}) \)
{[}\ref{we3}{]}. This geometrical dependence of the tensor matrix element is
separated out from the dynamics by using the Wigner-Eckart theorem. As a result,
the dynamics of the DPI process resides in the reduced matrix elements and the
geometric dependence is contained in the angular part.

We define

\begin{equation}
\label{13a}
\zeta =\sqrt{2J_{c}+1}\sqrt{2k_{1}+1}\sqrt{2J_{b}+1}\sqrt{2J^{\prime }_{b}+1}\sqrt{2k_{2}+1}
\end{equation}
 and

\begin{equation}
\label{13b}
\xi =\left\langle J_{b}\right\Vert j_{1}\left\Vert J_{a}\right\rangle \left\langle J_{b}\right\Vert j^{\prime }_{1}\left\Vert J_{a}\right\rangle ^{\star }\left\langle J_{c}\right\Vert j_{2}\left\Vert J_{b}\right\rangle \left\langle J_{c}\right\Vert j^{\prime }_{2}\left\Vert J_{b}\right\rangle ^{\star }.
\end{equation}
 Then the expectation value of the efficiency operator in Eq.(\ref{12}) becomes

\begin{equation}
\label{13c}
\begin{array}{cc}
\overline{\varepsilon } & \sim \sum (-1)^{J_{a}}\zeta \xi \left\{ \begin{array}{ccc}
J_{c} & j_{2} & J_{b}\\
J^{\prime }_{c} & j^{\prime }_{2} & J^{\prime }_{b}\\
0 & k_{2} & k_{2}
\end{array}\right\} \left\{ \begin{array}{ccc}
J_{b} & j_{1} & J_{a}\\
J^{\prime }_{b} & j^{\prime }_{1} & J^{\prime }_{a}\\
k_{2} & k_{1} & k
\end{array}\right\} \\
 & \times C_{\kappa _{2}\kappa _{1}\kappa }^{k_{2}k_{1}k}z_{k_{1}}(1)z_{k_{2}}(2)\rho ^{\gamma }_{k\kappa }(1,1)c_{k_{1}\kappa ^{\prime }_{1}}(j_{1}j^{\prime }_{1})\\
 & \times c_{k_{2}\kappa ^{\prime }_{2}}(j_{2}j^{\prime }_{2})C_{\kappa _{1}\kappa _{2}\kappa }^{k_{1}k_{2}k}C_{\kappa ^{\prime }_{1}\kappa ^{\prime }_{2}\kappa ^{\prime }}^{k_{1}k_{2}k}D_{\kappa \kappa ^{\prime }}^{k}(\Re )\, .
\end{array}
\end{equation}

\subsection{Attenuation corresponding to polarization sensitivity of a detector}

The electron detector may or may not be sensitive to the spin state of the incoming
electron. The attenuation of the signal due to the detector will depend on this
sensitivity. The factor \( c_{k_{i}\kappa _{i}}(j_{i}j^{\prime }_{i}) \), \( (i=1,2) \)
describes this property {[}\ref{we3}{]}. We shall now consider two different
cases.

\subsubsection{Case-I : Detectors insensitive to electron polarization}

If the detectors(CMAs) are insensitive to the spin polarization of electrons
then the projection \( \kappa _{i} \) of the \( k_{i} \)th component of the
angular momentum is effectively zero, i.e. the electrons are emitted symmetrically
with respect to the axis of the detector. Hence the attenuation factor can be
written as

\begin{equation}
\label{14}
c_{k_{i}0}(j_{i}j^{\prime }_{i})=\frac{\sqrt{2j_{i}+1}\sqrt{2j^{\prime }_{i}+1}}{4\pi }(-1)^{j_{i}-\frac{1}{2}+k_{i}}C_{\frac{1}{2}-\frac{1}{2}0}^{j_{i}j^{\prime }_{i}k_{i}}\, .
\end{equation}

\subsubsection{Case-II : Detectors sensitive to electron polarization}

In reference {[}\ref{we2}{]} we defined \( c_{k_{i}\kappa _{i}}(j_{i}j^{\prime }_{i}) \)
as the attenuation factor due to the change in the state of polarization of
an electron caused by the detector. When the detectors are insensitive to electron
polarization, one takes the average over the electron spin and its projection.
Now consider the case where the detectors are sensitive to electron polarization.
In this case the spin sensitivity of the detectors is described by a tensor
of the form \( c_{k_{s_{i}}\kappa _{s_{i}}}(s_{i}s_{i}) \). The attenuation
factor then turns out to be

\begin{equation}
\label{15}
\begin{array}{cc}
c_{k_{i}\kappa _{i}}(j_{i}j^{\prime }_{i})= & c_{k_{l_{i}}0}(l_{i}l^{\prime }_{i})c_{k_{s_{i}}\kappa _{s_{i}}}(s_{i}s_{i})\sqrt{2k_{l_{i}}+1}\sqrt{2k_{s_{i}}+1}\\
 & \times \sqrt{2j_{i}+1}\sqrt{2j_{i}^{\prime }+1}\left\{ \begin{array}{ccc}
l_{i} & l^{\prime }_{i} & k_{l_{i}}\\
s_{i} & s_{i} & k_{s_{i}}\\
j_{i} & j^{\prime }_{i} & k_{i}
\end{array}\right\} C_{0\kappa _{i}\kappa _{i}}^{k_{l_{i}}k_{s_{i}}k_{i}}
\end{array}\, ,
\end{equation}
 where

\begin{equation}
\label{16}
c_{k_{l_{i}}0}(l_{i}l^{\prime }_{i})=\frac{\sqrt{2l_{i}+1}\sqrt{2l^{\prime }_{i}+1}}{4\pi }(-)^{l^{\prime }_{i}}C_{000}^{l_{i}l^{\prime }_{i}k_{l_{i}}}\, ,
\end{equation}
 and \( c_{k_{s_{i}}\kappa _{s_{i}}}(s_{i}s_{i}) \) can be expressed in terms
of the Stokes parameters describing the spin polarization of the electron to
be detected{[}\ref{schmidt}{]}. The factor \( c_{k_{s_{i}}\kappa _{s_{i}}}(s_{i}s_{i}) \)
picks out electrons with a particular spin projection and may be called a \emph{Stern-Gerlach
operator}. Its components are

\begin{equation}
\label{17}
\begin{array}{cc}
c_{00}=\frac{1}{\sqrt{2}} & c_{10}=\frac{S^{e}_{z}}{\sqrt{2}}\\
c_{11}=-(S^{e}_{x}-iS^{e}_{y}) & c_{1-1}=-(S^{e}_{x}+iS^{e}_{y})\, .
\end{array}
\end{equation}
 Here \( S^{e}_{x},S^{e}_{y} \) and \( S^{e}_{z} \) are Stokes parameters
describing the polarization of the electron. For polarization insensitive detectors
one has \( S_{x}^{e}=S_{y}^{e}=S_{z}^{e}=0 \), and the attenuation factor reduces
to Eq.(\ref{14}). 

The lifetime of the singly ionized state is very small. Depending on the photon
energy there may be a situation where it is impossible to differentiate between
the photo- and Auger electrons simply by energy analysis. Then, to distinguish
between the two electrons it is necessary to measure the electron spin, i.e.
their polarization. For spin analysis of the electrons we have to use a Stern-Gerlach
type experimental set-up. Here the factor \( c_{k_{s_{i}}\kappa _{s_{i}}}(s_{i}s_{i}) \)
serves exactly that purpose, i.e. picks out electrons with a particular spin
projection. This type of experiment is known as `energy- and angle-resolved
coincidence experiment' and is being done by Schmidt and his co-workers {[}\ref{schmidt2}{]}.

In general, for DPI of atoms using polarized photon of sufficient energy, one
can distinguish the photo- and the Auger electrons by differential energy analysis.
In that case determination of electron spin is meaningless. Then, if the spin
is unobserved, one can take the average over the spin projection. In that case
the projection \( \kappa _{i} \) of the \( k_{i} \)th component of the angular
momentum is zero and the attenuation factor \( c_{k_{i}\kappa _{i}}(j_{i}j^{\prime }_{i}) \)
turns out to be \( c_{k_{i}0}(j_{i}j^{\prime }_{i}) \).

\section{Calculation and results}

In reference {[}\ref{we2}{]} we treated DPI in the xenon atom due to unpolarized
light. In this paper we are concerned with the same xenon atom with the difference
that DPI occurs due to a polarized light source. A randomly oriented xenon atom
is irradiated with a polarized photon beam of energy \( 94.5\, eV \). As a
result, the xenon atom no longer remains randomly oriented but acquires the
polarization of the photon beam. This leads to photoionization in the \( 4d_{5/2} \)
shell followed by a subsequent \( N_{5}-O_{23}O_{23}\, ^{1}S_{0} \) Auger decay.
We use the dipole approximation, the letters \( e,f \) and \( g \) for the
three possible photoionization channels {[}\ref{we2}{]}. These are characterised
by \( e)4d_{5/2}\longrightarrow \varepsilon _{p}f_{7/2} \), \( f)4d_{5/2}\longrightarrow \varepsilon _{p}f_{5/2} \)
and \( g)4d_{5/2}\longrightarrow \varepsilon _{p}p_{3/2} \) respectively. And
the Auger transition is characterised by the wave \( \varepsilon _{A}d_{5/2} \).
The same selection rules for photoionization and Auger transitions hold good
as in the case of unpolarized light.

In experiments for measuring angular correlation one usually chooses detectors
which are insensitive to the spin polarization of electrons. In such a case
\( \kappa _{1}=\kappa _{2}=\kappa ^{\prime }_{1}=\kappa ^{\prime }_{2}=\kappa =\kappa ^{\prime }=0 \),
and \( D^{k}_{00}(\beta _{1}\theta \beta _{2})=P_{k}(\cos \theta ) \). Then
Eq. (\ref{13c}) becomes

\begin{equation}
\label{18}
\begin{array}{cc}
\overline{\varepsilon } & \sim \sum (-1)^{J_{a}}\zeta \xi \left\{ \begin{array}{ccc}
J_{c} & j_{2} & J_{b}\\
J^{\prime }_{c} & j^{\prime }_{2} & J^{\prime }_{b}\\
0 & k_{2} & k_{2}
\end{array}\right\} \left\{ \begin{array}{ccc}
J_{b} & j_{1} & J_{a}\\
J^{\prime }_{b} & j^{\prime }_{1} & J^{\prime }_{a}\\
k_{2} & k_{1} & k
\end{array}\right\} z_{k_{1}}(1)z_{k_{2}}(2)\rho ^{\gamma }_{k\kappa }(1,1)\\
 & \times C_{000}^{k_{2}k_{1}k}c_{k_{1}0}(j_{1}j^{\prime }_{1})c_{k_{2}0}(j_{2}j^{\prime }_{2})P_{k}(\cos \theta ).
\end{array}
\end{equation}
 The summation extends over \( k_{1},k_{2} \) and \( k \).

In the limiting case of unpolarized photons Eq.(\ref{18}) reduces to a simple
form. Using Eq. (\ref{14}) and some properties of \( 9-j \) symbols and Racah
coefficients {[}\ref{satchler}{]}, we get

\begin{equation}
\label{19}
\begin{array}{cc}
\overline{\varepsilon }= & \sum _{k}z_{k}(1)z_{k}(2)(-1)^{j_{1}+j_{2}}c_{k0}(j_{1}j^{\prime }_{1})c^{\star }_{k0}(j_{2}j^{\prime }_{2})\\
 & \times \left\langle J_{c}\right\Vert j_{1}\left\Vert J_{b}\right\rangle \left\langle J_{c}\right\Vert j^{\prime }_{1}\left\Vert J_{b}\right\rangle ^{\star }\left\langle J_{b}\right\Vert j_{2}\left\Vert J_{a}\right\rangle \left\langle J_{b}\right\Vert j^{\prime }_{2}\left\Vert J_{a}\right\rangle ^{\star }\\
 & \times w(J_{b}J^{\prime }_{b}j_{1}j^{\prime }_{1};kJ_{a})w(J_{b}J^{\prime }_{b}j_{2}j^{\prime }_{2};kJ_{c})P_{k}(\cos \theta )\, .
\end{array}
\end{equation}
 Note that this is identical with Eq.(25) of reference{[}\ref{we2}{]}, as it
should be.

Experiments on the xenon atom were carried out by Schmidt and his co-workers
using \( 94.5\, eV \) synchrotron radiation {[}\ref{schmidt}{]}. They used
a perpendicular plane geometry to describe the process. The collision frame
\( x,y,z \) is attached to the target where the \( z \) axis coincides with
the direction of the photon beam. The arbitrary polarization of the incident
beam from the synchrotron is described by the Stokes parameters \( S_{1},S_{2} \)
and \( S_{3} \). Both \( S_{1} \) and \( S_{2} \) refer to the same quantity,
but with differently oriented axes. One can make \( S_{2}=0 \) by choosing
the \( x \) axis of the collision frame to coincide with the direction of maximum
linear polarization, i.e. the major axis of the polarization ellipse. To compare
our results with experimental values we use the same polar and azimuthal angles
in the perpendicular plane geometry. Fig.1 shows the perpendicular plane geometry
described above, \( e_{1} \) and \( e_{2} \) being the directions of emission
of the photo- and the Auger electron respectively. \( \theta  \) is the angle
between their directions of emission. We have calculated the theoretical value
of the angular correlation function for two different cases.\\
(i) The photo-electron is observed in a fixed direction and the Auger electron
spectrometer is turned around to get the angular distribution of the Auger electrons
with respect to the photo-electron. Here the maximum allowed value of \( k \)
is \( 2j_{2} \). \\
(ii)The second one is the complementary case, i.e. the Auger electron is observed
in a fixed direction and the photo-electron spectrometer is turned around to
get the angular distribution of the photo-electrons with respect to the Auger
electron. Here the maximum allowed value of \( k \) is \( 2j_{1,\max } \),
\( j_{1,\max } \) is the maximum value of \( j_{1} \) for the possible photoionization
channels.

The value of \( k \) gives the highest order of the Legendre polynomials occurring
in the correlation function. Interchannel interaction of the different photo-electron
channels contributes to the angular correlation pattern by introducing the different
terms, however, the total intensity remain unchanged. This inter channel interaction
is treated as it was in reference {[}\ref{we2}{]}.

As in reference {[}\ref{we2}{]} we have defined the angular correlation function
to be the angular part of the expectation value of the efficiency operator.
Solid lines represents the theoretically calculated plot and the dots represents
the experimental plot {[}\ref{schmidt}{]}. For a linearly polarized incident
photon beam the angular correlation function for our case turns out to be

i) Case 1: \( S_{1}=1,S_{2}=0,S_{3} \) unknown. The photo-electron is observed
in a fixed direction and the Auger electron spectrometer is turned around to
get the angular distribution of the Auger electron with respect to the photo-electron.

\begin{equation}
\label{20}
W(\theta )\sim 1+1.314P_{2}(\cos \theta )+1.100P_{4}(\cos \theta )\, .
\end{equation}

ii) Case 2: \( S_{1}=1,S_{2}=0,S_{3} \) unknown. The Auger electron is observed
in a fixed direction and the photo-electron spectrometer is turned around to
get the angular distribution of the photo-electron with respect to the Auger
electron.

\begin{equation}
\label{21}
W(\theta )\sim 1+0.817P_{2}(\cos \theta )+0.602P_{4}(\cos \theta )+0.570P_{6}(\cos \theta )\, .
\end{equation}

In both the cases one of the electron spectrometers is kept fixed along the
direction of the electric field vector (x-axis). The index \( k \) in the general
theoretical expression for \( \overline{\varepsilon } \) depends on the angular
momenta of the emitted electrons. Hence, the structure of the angular correlation
pattern depends on this index. If higher order angular momenta are involved
the angular correlation pattern has more structure. This is clear from the figures
2 and 3. Since the distribution of the photo-electron with respect to the fixed
Auger electron direction involves higher order angular momenta, the angular
correlation patterns has more structure.

\vspace{0.3cm}
{\par\centering \includegraphics{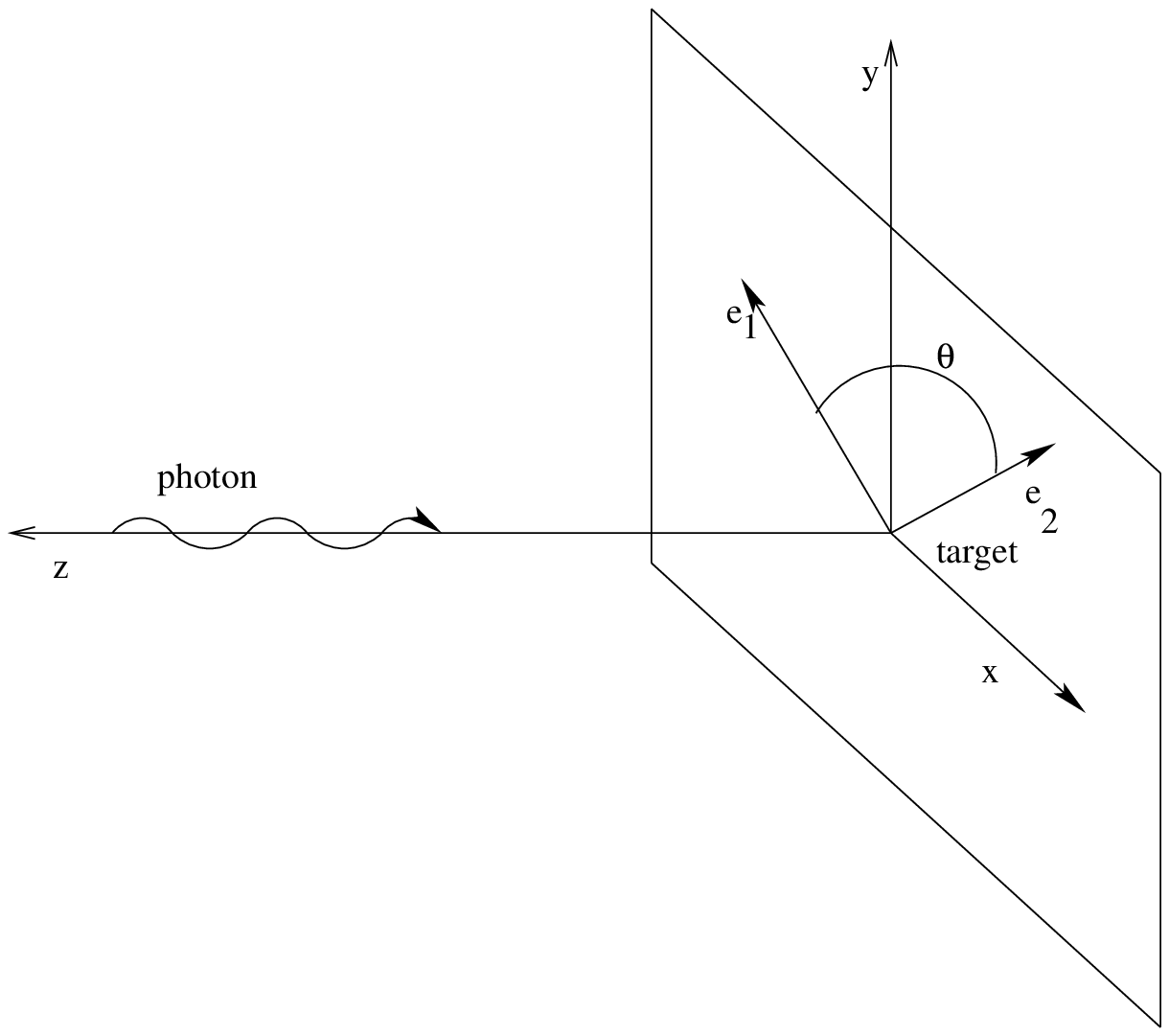} \par}
\vspace{0.3cm}

{\par\centering Figure 1 : Perpendicular plane geometry used in the experiments
by Schmidt and his co-workers\par}

\vspace{0.3cm}
{\par\centering \includegraphics{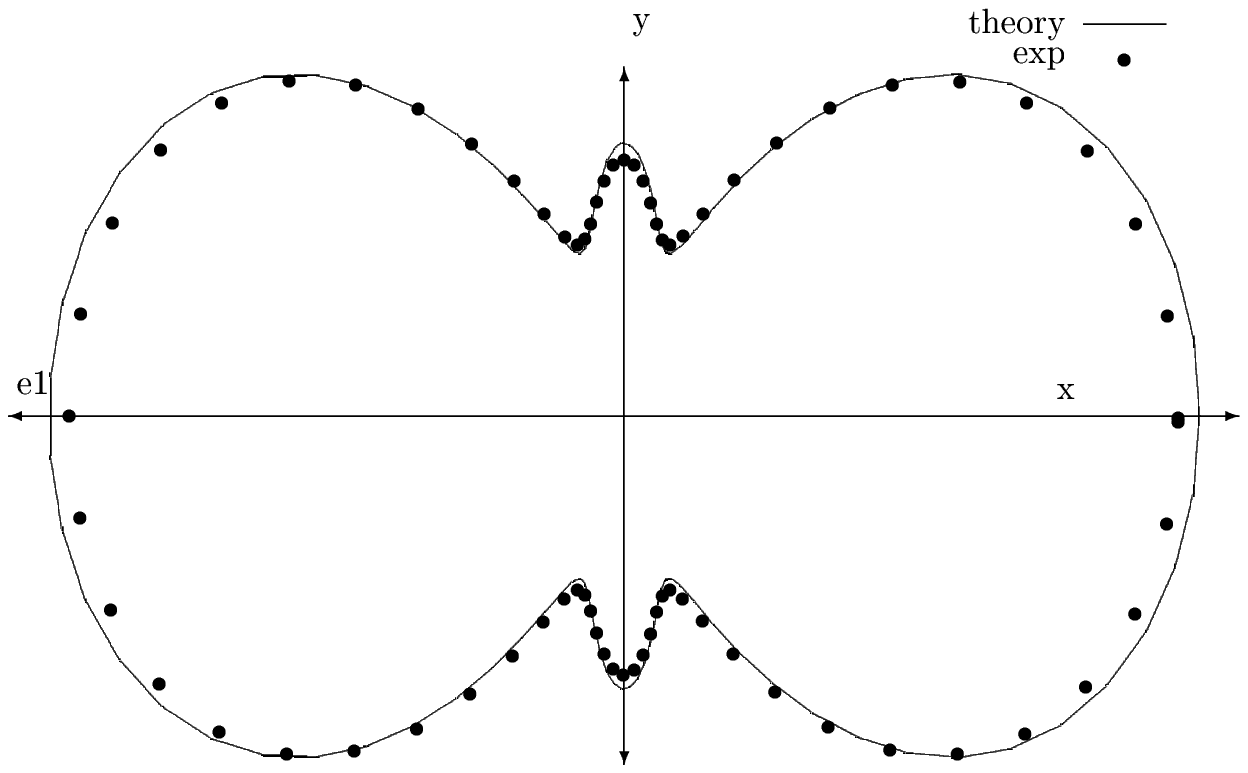} \par}
\vspace{0.3cm}

{\par\centering Figure 2. Angular correlation pattern for xenon due to a linearly
polarized photon beam(\( S_{1}=1,S_{2}=0,S_{3} \) unknown) of \( 94.5\, eV \)
(\( 4d_{5/2} \) photoionization followed by \( N_{5}-O_{23}O_{23}\, ^{1}S_{0} \)
Auger decay). The photo-electron is observed in a fixed direction{[}\ref{schmidt}{]}.\par}

\vspace{0.3cm}
{\par\centering \includegraphics{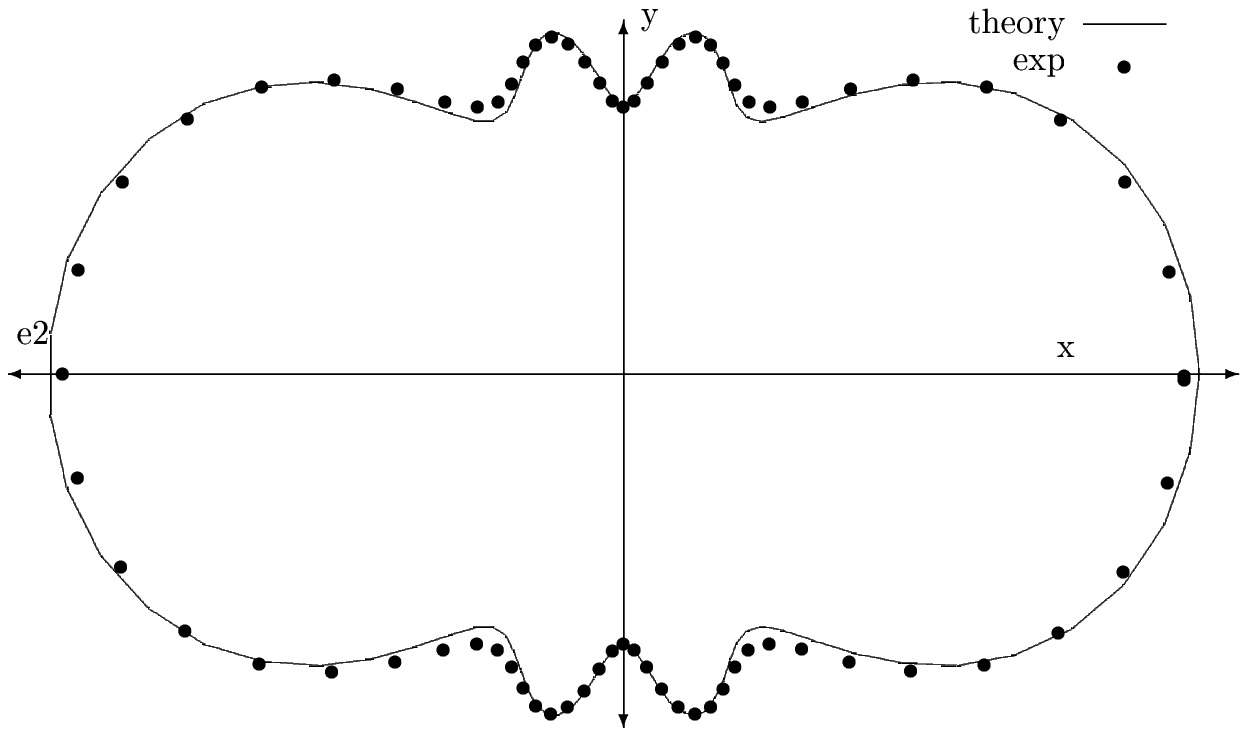} \par}
\vspace{0.3cm}

{\par\centering Figure 3. Angular correlation pattern for xenon due to a linearly
polarized photon beam(\( S_{1}=1,S_{2}=0,S_{3} \) unknown) of \( 94.5\, eV \)
(\( 4d_{5/2} \) photoionization followed by \( N_{5}-O_{23}O_{23}\, ^{1}S_{0} \)
Auger decay). The Auger electron is observed in a fixed direction{[}\ref{schmidt}{]}.\par}

\textbf{Acknowledgment:} One of the author (CS) would like to acknowledge the
financial support provided by the University Grants Commission of India.

\end{document}